\documentstyle[pre,aps]{revtex}

\newcommand{\beq}{\begin{equation}}
\newcommand{\eeq}{\end{equation}}
\newcommand{\bea}{\begin{eqnarray}}
\newcommand{\eea}{\end{eqnarray}}
\newcommand{\req}[1]{Eq.~(\ref{#1})}
\begin{document}
\draft
\title{Thermodynamic functions of harmonic Coulomb crystals}

\author{D. A. Baiko,  A. Y. Potekhin, and D. G. Yakovlev}
\address{Ioffe Physico-Technical Institute,
        194021 St.\ Petersburg, Russia}
\date{Received 20 April 2001; accepted 17 August 2001}
\maketitle

\begin{abstract}
Phonon frequency moments
and thermodynamic functions (electrostatic and vibrational 
parts of the free energy, internal energy, and heat capacity)
are calculated for bcc and fcc Coulomb crystals
in the harmonic
approximation with a fractional accuracy $\lesssim10^{-5}$. Temperature
dependence of thermodynamic functions is fitted by analytic formulas
with an accuracy of a few parts in $10^5$.
The static-lattice (Madelung) 
part of the free energy is calculated with an 
accuracy $\sim10^{-12}$. The Madelung constant and
frequency moments of hcp crystals are also computed.
\end{abstract}


% 05.70.Ce: Thermodynamic functions and EOS
% 5. Physics of gases, plasmas, electric discharges:
% 52. Physics of plasmas and el.discharges
% 52.25.-b: Plasma properties
% 52.25.Kn: Thermodynamics of plasmas
% 52.27.Gr: Strongly-coupled plasmas
% 95. Fundamental A&A
% 95.30.Qd: MHD and plasmas
% 97. Stars
% 97.20.Rp: Faint blue stars, WDs, degenerate stars...
% 97.60.Jd: Neutron stars

% **************************************************************
%                               TEXT BODY
% **************************************************************

%%  SECTION   **************************************************
\section{Introduction}
Coulomb crystals, introduced into theory by Wigner
\cite{Wigner}, have been studied by many authors.
A thorough discussion was given, e.g., in
Refs.\cite{Kugler,PH73}.
Ewald technique \cite{Ewald,BH54} has been used
to calculate shear constants \cite{Fuchs} and dispersion relations
 \cite{Clark} of such crystals.
The thermodynamics in the harmonic-lattice approximation
has been analyzed, e.g., in Refs.\cite{PH73,Carr,C93}.
Anharmonic corrections have been discussed in
Refs.\cite{Albers86,Dubin,FH}. Chabrier et al.\cite{CAD} 
have suggested an approximate analytic model of the harmonic 
Coulomb crystal, which is widely used in astrophysics
(e.g., Refs.\cite{C93,WD,WD1}). 
However, precise numerical calculations of the
thermodynamic functions, valid at any temperature $T$,
have not been published.

Here we report highly accurate calculations of
phonon spectra and frequency moments
of body-centered-cubic (bcc),
face-centered-cubic (fcc), and
hexagonal-close-packed (hcp)
one-component Coulomb lattices in the harmonic approximation.
We present also accurate calculations
of thermodynamic functions for bcc and fcc lattices
at any values of the quantum parameter $\theta=T_p/T$, where
$T_p=\hbar\omega_p/k_B$ is the ion plasma temperature and 
$\omega_p=\sqrt{4 \pi n_i Z^2 e^2/ M}$ is the ion plasma 
frequency ($n_i$, $M$, and $Ze$ being the ion number density, 
mass, and charge, respectively). The numerical results
are given in the easy-to-use form of tables and fitting formulas.

%%  SECTION   **************************************************
\section{Phonon spectrum and electrostatic energy}
\label{sect-spectrum}
Consider a crystal of identical ions
immersed in the uniform compensating background. 
The basic definitions are as follows (e.g., Ref.\ \cite{BH54}).
Let us take an arbitrary ion as the origin of a Cartesian
reference frame and specify the lattice basis
${\bf l}_1,{\bf l}_2,{\bf l}_3$
generating direct lattice vectors
${\bf l}(n_1,n_2,n_3) = n_1 {\bf l}_1+n_2 {\bf l}_2+n_3 {\bf l}_3$,
where $n_1,n_2,n_3$
are arbitrary integers.
The vectors ${\bf g}(n_1,n_2,n_3) =
n_1 {\bf g}_1+n_2 {\bf g}_2+n_3 {\bf g}_3$,
where ${\bf g}_i \cdot {\bf l}_j = 2\pi\delta_{ij}$, form
the reciprocal lattice. Consider also the primitive cell, the
parallelepiped $\{\nu_1 {\bf l}_1+\nu_2 {\bf l}_2+\nu_3 {\bf
l}_3 \}$ with $0 \le \nu_1, \nu_2, \nu_3 < 1 $.
Let $N_{\rm cell}$ be the number of ions in the
primitive cell enumerated with an index $k$. 
The choice of the vectors ${\bf l}_i$ is not unique,
and one can describe a given lattice using different $N_{\rm cell}$.
We will adopt the standard convention and choose
the primitive cell with the lowest $N_{\rm cell}$.
The bcc and fcc lattices are simple
(the lowest $N_{\rm cell}=1$), whereas for the hcp lattice
one has the lowest $N_{\rm cell}=2$.

Along with the primitive cell one usually considers 
the Wigner-Seitz (WS) cell, which is a polyhedron
with faces crossing the lattice vectors
at their midpoints at the right angle. The volume of the WS cell is equal
to that of the primitive cell, $N_{\rm cell}/n_i$. 
A convenient measure of interparticle spacing is the ion-sphere radius 
$a=(3/4 \pi n_i)^{1/3}$.
%%% If $N_{\rm cell}=1$, the volume of a sphere of radius $a$
%%% is equal to that of the WS cell. 
The WS cell of the reciprocal lattice 
is the first Brillouin zone (BZ);
its volume equals $V_{BZ}=(2\pi)^3 n_i$.

The frequencies $\omega_s$ and polarization
vectors ${\bf e}_s$ of lattice vibrations 
($s=1, \ldots, 3 N_{\rm cell}$ enumerates vibration modes)
at any point ${\bf q}$ of the BZ are determined
by (e.g., Ref.\ \cite{BH54}) 
\begin{equation}
     {\omega^2 \over \omega_p^2} \, e^{\alpha k}_s -
     \sum_{k'\beta} {\cal D}_{\alpha \beta}(k,k',{\bf q}) \,
     e^{\beta k'}_{s} = 0,
\label{secul}
\end{equation}
where the summation is over three 
Cartesian coordinates (Greek indices) and over
the ions in the primitive cell ($k'$); 
\begin{eqnarray}
    {\cal D}_{\alpha \beta}(k,k',{\bf q}) &=&
    {1 \over 3} \, \delta_{\alpha \beta} \delta_{kk'} -
    {a^3 \over 3} \left( 
    {\partial^2 \over \partial u_\alpha \partial u_\beta}
    {\sum_{\bf l}}'
    {{\rm e}^{-i {\bf q}\cdot {\bf l}} \over |{\bf r} - {\bf u}|}
    \right)_{{\bf u} \to 0}
\end{eqnarray}
is the dynamical matrix,
${\bf r} = {\bf l} + {\bf x}(k) - {\bf x}(k')$, and 
${\bf x}(k)$ specifies the ion position within the primitive cell.
The primed sum means that the term 
${\bf l}=0$ is excluded if $k=k'$.

The elements of the dynamical matrix
can be calculated using the Ewald technique of
theta-function transformations (e.g., Ref.\cite{BH54}),
which yields
\begin{eqnarray}
   && {\cal D}_{\alpha \beta}(k,k',{\bf q})
     =
    \frac{1}{3} \, \delta_{\alpha \beta} \delta_{kk'} -
     {4 \rho^3 a^3 \over 9 \sqrt{\pi}}
    \delta_{\alpha \beta} \delta_{kk'} - 
     {\sum_{\bf l}}'
    \left[ \rho^3 a^3 {
     4 \over 3 \sqrt{\pi}} {r_{\alpha} r_{\beta} \over r^2}
    e^{-\rho^2 r^2}  
\right.\nonumber\\ 
          && \left.
    + { \rho a^3 \,
      (3 r_{\alpha} r_{\beta} - \delta_{\alpha \beta} r^2) \over 3 r^4}
     \left( \frac{{\rm erfc}(\rho r)}{\rho r} + 
    {2 \over \sqrt{\pi}} e^{-\rho^2 r^2} \right)
      \right] e^{-i{\bf q}\cdot{\bf l}}
\nonumber\\
      && + { 1 \over N_{\rm cell}} \sum_{\bf g}
      \frac{(g_\alpha - q_\alpha) \, (g_\beta - q_\beta)}
        {\left| {\bf g} - {\bf q} \right|^2} \,
      \exp{\left( - {\left| {\bf g} - {\bf q} \right|^2 \over 4 \rho^2 }  
      -i \, ({\bf g} - {\bf q}) \cdot
      [{\bf x}(k) - {\bf x}(k')]   \right)}.
\label{calD}
\end{eqnarray}
The last sum is over all reciprocal lattice vectors,
and $\rho$ is a free parameter adjusted
to yield equally rapid convergence of direct and
reciprocal sums; a suitable choice is $\rho a \approx 2$.
Numerical calculations according to Eq.\ (\ref{calD}) become unstable 
at $qa \ll 1$ (near the BZ center). In this region 
we replace $D_{\alpha \beta}$ by an appropriate asymptote \cite{CK55},
whose coefficients have been
recalculated with an accuracy $\sim 10^{-8}$ using the Ewald technique.

The static-lattice binding energy
of a Coulomb lattice is
\beq
   E_0= K_M Z^2 e^2 / a,
\eeq
where the Madelung constant $K_M$ can
be written as:
\begin{eqnarray}
   K_M &=&   {a \over 2 N_{\rm cell}}
             \sum_{k',k} {\sum_{\bf l}}' 
           \frac{{\rm erfc}(\rho r)}{r}
              - \frac{3}{8 \rho^2 a^2} -
            {\rho a \over \sqrt{\pi}}
%%% \nonumber \\ &&
          + {3 \over 2 N^2_{\rm cell}}   
        \sum_{k',k} {\sum_{\bf g}}' \frac{1}{g^2 a^2}
             \exp \left(-{ {\bf g}^2 \over 4 \rho^2 }
             % \right)} e^{ 
            + i {\bf g}\cdot[{\bf x}(k) - {\bf x}(k')] \right).
\label{fuchs}
\end{eqnarray}
Previously $K_M$ was calculated, e.g., in Refs.\cite{Maradudin,BST}.
Our calculated values of $K_M$
for bcc, fcc, and hcp crystals are given in Table~\ref{tab-lattice}.

%%  SECTION   **************************************************
\section{BZ integration and frequency
moments}
\label{sect-integration}
In many physical problems, one needs to average
functions $f(\omega)$ over phonon branches and wave vectors:
\begin{equation}
   \langle f \rangle = { 1 \over 3 N_{\rm cell}} \,
   \sum_s \bar{f}, \quad \bar{f}=  
{1 \over V_{BZ}} 
   \int_{\rm BZ} f({\bf q}){\rm\,d} {\bf q} ,
\label{intIR}
\end{equation}
where $f({\bf q})\equiv f(\omega_s({\bf q}))$.
In Eq.\ (\ref{intIR}) we will use the Holas integration method
considered in Ref.\ \cite{AG81} for the bcc lattice:
\begin{equation}
   \bar{f} =
         \int_0^1 {\rm d} \xi  \int_0^1{\rm d} \eta 
         \int_0^1{\rm d} \zeta   \, \eta \, \xi ^2
         {\cal F}\{f\},
\label{bccint}
\end{equation}
where $\xi $, $\eta $, and $\zeta $ are appropriate
BZ coordinates.
For the bcc crystal, ${\cal F}\{f\}=6f({\bf q})$, with
${\bf q} \equiv  (q_x,q_y,q_z) = 
       ( 2 - \eta  ,  \eta  , \eta  \zeta  ) \pi \xi /a_l$,
and the lattice constant $a_l$ is given by $n_i a_l^3 = 2$.
We calculate the integrals in Eq.\ (\ref{bccint}) by
the Gauss method involving
the nodes of the Jacobi polynomials $P_n^{(0,0)}$.
The integral over $\eta $ is alternatively treated
by the generalized Gauss scheme with weight function $\eta $,
which involves the nodes of $P_n^{(1,0)}$.

This approach can be also developed
for the fcc and hcp lattices. In both cases 
we come again to Eq.\ (\ref{bccint}), but with 
different ${\cal F}\{f\}$. For the fcc lattice, we have
$
    {\cal F}\{f\} = \frac32 [ \frac32 f({\bf q}_1)+
                            \frac32 f({\bf q}_2)+
                            f({\bf q}_3)],
$
where ${\bf q}_i= {\bf Q}_i \, \pi \xi  /(2 a_l)$,
        ${\bf Q}_1= (2 + \eta  + \eta  \zeta ,
        2 + \eta  - \eta  \zeta , 2 - 2\eta )$,
        ${\bf Q}_2 =  (2 + 2 \eta ,
        2 - \eta  + \eta  \zeta ,  2 - \eta  - \eta  \zeta )$,
        ${\bf Q}_3 = (4, \eta  + \eta  \zeta ,
        \eta  - \eta  \zeta )$, and $n_i a_l^3 =4$.

For the hcp lattice,
$
    {\cal F}\{f\} = 2 f({\bf q}_1)/\eta  +
                      2 f({\bf q}_2)$,
where ${\bf q}_i= {\bf Q}_i \, 2 \pi \xi  /(3 a_l)$,
        ${\bf Q}_1 = (\sqrt{3},
          \zeta , \frac32 \eta  / \sigma)$,
        ${\bf Q}_2 = (\eta \sqrt{3},
           \eta  \zeta , \frac32/ \sigma)$, 
       and $n_i a_l^3 =\sqrt{2}$.
Here, $\sigma = \sqrt{8/3}$ is twice the ratio
of the distance 
between the hcp lattice
planes to the distance between neighbors
within one plane.

Phonon frequency moments $\langle(\omega/\omega_p)^n\rangle$ and 
the average $\langle\ln(\omega/\omega_p)\rangle$,
obtained by this method, are 
given in Table~\ref{tab-lattice}.
We remind that 
$\langle(\omega/\omega_p)^2\rangle=\frac13$, according to the Kohn rule
(e.g., Ref.\ \cite{Maradudin}).
The accuracy of the data in Table~\ref{tab-lattice}
corresponds to the number of
digits shown; it is the same or higher than the accuracy of 
the previous results (e.g., Refs.\ \cite{Dubin,Maradudin,AG81,Nagai}), 
except only the value of $\langle \omega/\omega_p\rangle$ 
for the hcp lattice, calculated more accurately in Ref.\ \cite{Nagai}
($\langle \omega/\omega_p \rangle_{\rm hcp}=0.5133368$).

%%  SECTION   **************************************************
\section{Thermodynamic functions} 
\label{sect-res}
Free energy $F$ of a 
harmonic Coulomb crystal consists of the static-lattice
contribution $E_0$, contribution from zero-point ion vibrations,
$\frac32\,N\hbar \langle \omega \rangle$,
and thermal free energy in the harmonic
lattice approximation, $F_{\rm th}$.
Accordingly, the reduced free energy
$f\equiv F/(Nk_B T)$ is
\beq
f = K_M\,\Gamma + 1.5 \,\langle {{w}}\rangle+ f_{\rm th},
\label{fsum}
\eeq
where
$
 f_{\rm th}(\theta) =3\left\langle \ln\left(1-e^{-{{w}}}\right)
        \right\rangle,
$
and
\[
\Gamma={(Ze)^2\over a k_B T},
\quad
{{w}} = {\hbar\omega\over k_B T} = \theta\,{\omega\over \omega_p}.
\]
Thus, the reduced internal energy
$ u \equiv U/(Nk_B T) = -\partial f /\partial\ln T$ is
\beq
u =
K_M\,\Gamma + 1.5\,\langle {{w}}\rangle + 
          u_{\rm th},
\eeq
where
\beq
      u_{\rm th}(\theta) = {{\rm d} f_{\rm th}\over {\rm d} \ln \theta}
         = 3\left\langle{{{w}}\over e^{{w}}-1}\right\rangle.
\label{u-HL}
\eeq
The harmonic constituent of the reduced heat capacity, 
$
c_V = (Nk_B)^{-1}\,{\partial U/\partial T}=
      u + {\partial u/\partial\ln T},
$
is
\beq
   c_{V}(\theta) = u_{\rm th} -{{\rm d} u_{\rm th} \over {\rm d} \ln\theta}
= 3\left\langle {{{w}}^2\,e^{-{{w}}}\over(1-e^{-{{w}}})^2}\right\rangle .
\label{cV-HL}
\eeq

Using the results of Secs.\ \ref{sect-spectrum} and
\ref{sect-integration},
we have calculated 
$f_{\rm th}(\theta)$, $u_{\rm th}(\theta)$, and $c_V(\theta)$
for bcc and fcc crystals as corresponding BZ averages.
The mean numerical error is estimated as
$\sim 10^{-6}$, and it is a few times larger
at $\theta \gg 1$. 
Let us discuss possible analytic approximations. 
The model of Chabrier et al.\ \cite{CAD} assumes the linear dispersion law
for two acoustic (Debye-type) modes, $\omega_\perp=\alpha\omega_p\,q/q_B$,
and an optical (Einstein-type) mode,
$\omega_\| = \gamma \omega_p$.
The known phonon spectrum moments of a Coulomb crystal 
are approximately reproduced with the 
choice $\alpha\approx0.4$, $\gamma\approx0.9$.
In this model,
\begin{equation}
   f_{\rm th} = 2\,\ln\left(1-e^{-\alpha\theta}\right) +
                 \ln\left(1-e^{-\gamma\theta}\right)
                - {\textstyle\frac23} \,D_3(\alpha\theta),
\label{fCha}
\end{equation}
where 
$
D_3(z) \equiv(3/ z^3)\int_0^z{t^3/(e^t-1)}{\rm\,d}t\,
$
is the Debye function. 
This model reproduces
numerical values of $f_{\rm th}$, $u_{\rm th}$, and $c_V$
with an accuracy of $\sim10\%$.

A heuristic generalization of \req{fCha} provides
a convenient fitting formula to $f_{\rm th}$.
Introducing three logarithmic terms (according to three phonon modes) 
and replacing $D_3$
by an arbitrary rational-polynomial function possessing
the correct asymptote $\propto\theta^{-3}$
%(1+O(\theta^{-2}))$ 
at large $\theta$,
we obtain:
\bea
   f_{\rm th} = \sum_{n=1}^3\ln\left(1-e^{-\alpha_n\theta}\right) -
  {A(\theta)\over B(\theta)},
\label{HLfit}
\eea
where
\beq
 A(\theta)= \sum_{n=0}^8 a_n\,\theta^n,
\qquad
B(\theta)= \sum_{n=0}^7 b_n\,\theta^n + \alpha_6\,a_6\,\theta^9+\alpha_8\,a_8\theta^{11},
\label{ABpolynom}
\eeq
and the parameters $\alpha_n$, $a_n$, and $b_n$ are given 
in Table~\ref{tab-HLfit}.

Calculation of the harmonic
thermal energy and heat capacity from \req{HLfit}
using Eqs.\ (\ref{u-HL}) and (\ref{cV-HL}) yields:
\bea
   u_{\rm th} & = & \sum_{n=1}^3 {\alpha_n\,\theta\over e^{\alpha_n\,\theta}-1}
        -\theta\,{A'(\theta)\,B(\theta) - A(\theta)\,B'(\theta) \over B^2(\theta)},
\label{uHLfit}
\\
    c_V & = & \sum_{n=1}^3 {\alpha_n^2\,\theta^2\over
        (e^{\alpha_n\,\theta/2} - e^{-\alpha_n\,\theta/2})^2}
    + \theta^2\,{A''\,B^2 - 2A'\,B'\,B + 2\,A\,(B')^2 - A\,B\,B''\over B^3},
\label{cHLfit}
\eea
where the first and second derivatives $A'$, $A''$, $B'$, and $B''$
are readily obtained from \req{ABpolynom}.

The approximations (\ref{HLfit}), (\ref{uHLfit}), and (\ref{cHLfit})
have a fractional accuracy within $5 \times10^{-6}$,
$2\times10^{-5}$, and $5\times10^{-5}$, respectively.

In the classical limit $\theta\to0$,
the exact expansion of $f_{\rm th}$ is
\beq
  f_{\rm th} = 3\,\ln\theta + 
     3\,\left\langle\ln\left({\omega\over\omega_p}\right)\right\rangle 
      -\frac32\, \left\langle { \omega \over \omega_p } \right\rangle
      \,\theta + \frac{1}{24}\theta^2 + \ldots
\label{classic}
\eeq
Note that the term
$-\frac32 \, \langle \omega / \omega_p \rangle   \,\theta$
cancels the zero-point energy
in \req{fsum}.
Our fit (\ref{HLfit}) reproduces the logarithmic, constant, and linear
terms of Eq.\ (\ref{classic})  exactly (by construction),
whereas the last (quadratic) term is reproduced with the relative accuracy
of $5\times10^{-5}$ and $10^{-6}$ for bcc and fcc lattices, respectively.
Although we do not present calculations of the thermal thermodynamic
functions for hcp crystals, our analysis reveals that they do not
deviate from the functions for fcc crystals by more than
a few parts in $10^3$. 

Our results can be used in any applications which require
a fast and accurate evaluation of the thermodynamic functions
of the Coulomb crystals.

\begin{acknowledgements}
This work has been partly
supported by
RFBR Grant No.\ 99-02-18099.
A.P.\ thanks the theoretical astrophysics group at the Ecole Normale
Sup\'erieure de Lyon and the Department of Energy's Institute for
Nuclear Theory at the University of Washington for their hospitality and
partial support during the completion of this work.
\end{acknowledgements}

%%%%%%%%%%%%%%%%%%%%%%%%%%%%%%%%%%%%%%%%%%%%%%%%%%%%%%%%%%%%

%%%%%%%%%%%%%%%%%%%%%%%%%%%%%%%%%%%%%%%%%%%%%%%%%%%%%%%%%%%%

\begin{table}
\caption{Parameters of Coulomb crystals.}
\label{tab-lattice}
\begin{tabular}{ccccccc}
 lattice type & $K_M$ & $\langle (\omega/\omega_p)^{-2} \rangle$ &
                  $\langle (\omega/\omega_p)^{-1} \rangle$ & 
                  $\langle (\omega/\omega_p) \rangle$ & 
                  $\langle (\omega/\omega_p)^3 \rangle$ & 
  $\langle \ln{(\omega/\omega_p)} \rangle$ \\
\noalign{\smallskip}
\hline
\noalign{\smallskip}
 bcc & $-0.895\,929\,255\,682$ & 12.972 & $2.798\,55$ & $0.511\,3875$
       & $0.250\,31$ & $- 0.831\,298$ \\
 fcc & $-0.895\,873\,615\,195$ & 12.143 & $2.719\,82$ & $0.513\,1940$
       & $0.249\,84$ & $- 0.817\,908$ \\
 hcp & $-0.895\,838\,120\,459$ & 12.015 & $2.7026\phantom{\,0}$
       & $0.513\,33\phantom{00}$
       & $0.24984$ & $- 0.815\,97\phantom{0}$ \\
\end{tabular}
\end{table}

\begin{table}
\begin{center}
\caption{Parameters of the analytic approximation 
(\protect\ref{HLfit}) to the thermal free energies of
bcc and fcc Coulomb lattices. Powers of 10 are given in square brackets.}
\label{tab-HLfit}
\begin{tabular}{r|ccc|ccc}
 & \multicolumn{3}{c|}{bcc lattice} & \multicolumn{3}{c}{fcc lattice}\\
$n$     & $\alpha_n$ & $a_n$ & $b_n$ & $\alpha_n$ & $a_n$ & $b_n$ \\
\noalign{\smallskip}
\hline
\noalign{\smallskip}
0  & $-$      & 1               & 261.66    & $-$      & 1          & 303.20  \\
1  & 0.932446 & $0.1839$        & 0         & 0.916707 & 0          & 0      \\
2  & 0.334547 & 0.593586        & 7.07997   & 0.365284 & 0.532535   & 7.7255 \\
3  & 0.265764 & $5.4814\,[-3]$  & 0         & 0.257591 & 0          & 0      \\
4  & $-$ & $5.01813\,[-4]$      & 0.0409484 & $-$ & $3.76545\,[-4]$ & 0.0439597\\
5  & $-$ & 0              & $3.97355\,[-4]\,$ & $-$ & 0          & $1.14295\,[-4]$\\
6  & $4.757014\,[-3]$ & $3.9247\,[-7]$ & $5.11148\,[-5]$ & 
             $4.92387\,[-3]$ & $2.63013\,[-7]$ & $5.63434\,[-5]\,$\\
7  & $-$ & 0              & $2.19749\,[-6]\,$ & $-$ & 0               & $1.36488\,[-6]$\\
8  & $4.7770935\,[-3]$ & $5.8356\,[-11]$ &  $-$           & 
             $4.37506\,[-3]$ & $6.6318\,[-11]$ &  $-$       \\
\end{tabular}
\end{center}
\end{table}
\end{document}